# Uniformity of the quiet solar disk: 3130 – 46700 Å

W. Livingston, E. Galayda, and R. Milkey
National Solar Observatory, P.O. Box 26732, Tucson, AZ 85726, USA

## Abstract

Taking advantage of the absence of solar activity in the recent 2008-9 epoch (no spots, few faculae), we have made equatorial and meridian disk scans in continua from 3129 Å to 46700 Å. Averaging 20 scans at each wavelength to suppress granulation, which takes a total of 35 minutes, we achieve a system noise level of 0.01%. We believe this noise level is a record low, not because of instrument improvements, but simply because of observing procedures and the cooperation of the Sun and sky. The observed solar fluctuations significantly exceed the noise and range from 0.3% at 3130 Å, 0.05% at 34000 Å, to 0.06% at 46700 Å near disk center. These fluctuations (corresponding to about 3 K) presumably arise from the incomplete averaging of granulation. Standard solar models for limb darkening fit the data for true continuum regions reasonably well. No significant differences are seen between scan directions (E-W, N-S). Perhaps our results can serve as a template for exoplanet detection by the transit of quiet G2V-like stars.

## 1. Introduction

The existence of durable equipment continues to foster the long-term pursuit of limb-darkening work at the National Solar Observatory McMath-Pierce Telescope. The present paper may be considered an update of Livingston and Wallace (2003) with expanded spectral coverage into the IR at 46700 Å. In that paper we emphasized that the basal quiet atmosphere is observationally unaffected by the magnetic cycle. Another work of comparable low noise and spectral coverage (10000-40000 Å) is that of Koutchmy et al. (1977).

The uniformity of the solar disk is marred by wavelength dependent limb darkening, sunspots, faculae, and atmospheric "seeing" effects. During the 2008-9 epoch the Sun was exceptionally quiet with frequent spotless days and few faculae. Confining observations to cloud-free days, deleterious granular change and seeing may be suppressed by averaging.

Traditional limb darkening observations are made by turning off the telescope drive and allowing the solar image to drift across the spectrometer entrance slit. Except for short intervals when the position angle of the heliographic poles is near zero (i.e. around 7 July and 6 Jan), this means neither the exact equator or meridian is sampled. The spectrometer is set in the continuum (Pierce and Slaughter, 1977). This procedure has the advantage of constant air mass during the observation. It precludes meridian scans, however. It also means that in the past (Petro *et al*.1984, Neckel and Labs, 1994, and others), limb darkening has been sampled under varying degrees of activity. The exception was the work of Koutchmy et al. (1977) who scanned in heliographic coordinates near a time of minimum activity.

We should mention that pioneering limb darkening observations in the infrared began with P. Lena (1970) who used a Frank Low gallium-doped germanium bolometer at 5, 10, and 20 microns. He achieved a S/N of ~100 and a spatial resolution of 2-4 arc-sec. Noyes *et al.* (1967) also explored the Sun out to 1.2 mm where limb-brightening was detected. Others made observations at 1-3.7 microns using PbS cells (Peyturaux, 1955; Wöhl, 1975). Pierce *et al.*, 1977, used InSb to observe out to 2.4 microns, finding disk center fluctuations of $\sigma = 0.5\%$.

In the data to be presented, we are favored by an exceptionally quiet solar disk devoid of sunspots and free of almost any faculae. Activity in 2008 was at a 75 year low (Norton and Gallagher, 2009; see their Figure 6). To better sample the entire solar disk we have chosen to scan the image, so that we can examine the geographic meridian on a par with the geographic equatorial zone. Multiple consecutive scans, 20 in number, are made over a time interval as detailed below. These are averaged together to suppress the effects of granulation and p-mode oscillations. Residual solar structure, probably from exceptionally long-lived granulation patterns, remains however (Muller, 1988).

The wavelengths studied are (approximate) continuum positions at 3129*, 3883*, 4015, 5394, 8558, 15648* (the opacity minimum), 34100, and 46680* Å.

We investigate noise as a separate issue at wavelengths marked * above. We conclude that "system noise", in the 20 designated noise sets, is less than 0.01%. As will be seen, this is well below the observed disk fluctuations.

Finally, model-atmosphere limb darkening is compared to our observations out to a limb distance of 0.7 $R_\odot$. No attempt is made herein to study or quantify the extreme limb. This work is best done at eclipses or from satellites. Our data are posted at an ftp site: ftp://diglib.nso.edu/pub/wcl/limb.dark.scan.data.

## 2. Observations

The procedures employed are the same as in Livingston and Sheeley (2007), a study of spectrum-line variations in limb darkening data, except the Sun is now quiet, the wavelengths are at continuum positions, and each observation consists of an average of 20 scans of 4096 points. Each pair of observations ( *i.e.* N to S; S to N) is followed by the incremental re-positioning of the solar image "start position" by fixed limb guiders. The actual scan motion is "open loop", *i.e.* without guiders; solar rotation is not taken into account as it is unimportant in the continuum.

Recording is by the PHOTOMETRY program with time per sample about 1.4 msec, the number of samples per point set to 16, the number of points to 4096, and the number of scans 1 (meaning a pair). Entrance slit width is

0.5×10 mm, exit slit width is 0.2×10 mm, and the spectrograph is in "single pass". Taking into account an image scale of 2.37 arc-sec mm$^{-1}$, we average over a number of granules along the slit (1.19×23.7 arc-sec) even though the sample time corresponds to 0.53 arc-sec perpendicular to the slit.

Depending on the wavelength, grating order selection is by a colored glass filter or a prism pre-disperser (in the UV). The signal is from a quartz envelope photomultiplier tube or a cryogenically cooled anti-refection coated InSb diode.

Although we refer to observations as "equatorial" or "meridian", the scan directions are in sky coordinates (geographic coordinates), and are thus offset by the P-angle for the date, as given in Table 2. (We were aware that it would be preferable to scan in heliographic coordinates, but this was not possible because of available equipment. Besides, most past ground-based center-to-limb work has similarly been done in sky coordinates.)

The raw limb darkening data must be adjusted later, in the reduction process, to take into account minor (but significant) image starting position errors. (In other words the guider image reset function proves imperfect).

Each scan is preceded and followed by a shutter closure under the entrance slit. This serves to subtract dark current and amplifier detector offset and is a part of the PHOTOMETRY observing program. Each scan also begins and ends about two arc-minutes off the disk, so that sky scattered light is measured.

System noise was determined as follows. The 82 cm solar image is centered on the spectrograph and a 60 mm diameter quartz lens is centered in this beam, 3.5 m overhead, to form an image of the heliostat on the slit. This means we average over an area of about two arc-minutes on the disk, or a large number granules (we are uncertain how many). Unlike the limb darkening scans, the image is fixed. A single scan takes 1.5 minutes of time. The 20 scan sequence runs automatically, the data is written to disk, restarts, and takes an overall 30 minutes. These resulting records are averaged together later by a program called DECOMP, and a linear fit is made to the data which is normalized to 1.0. A typical noise record is seen in Figure 1. and summary of the noise record results is given in Table 1.

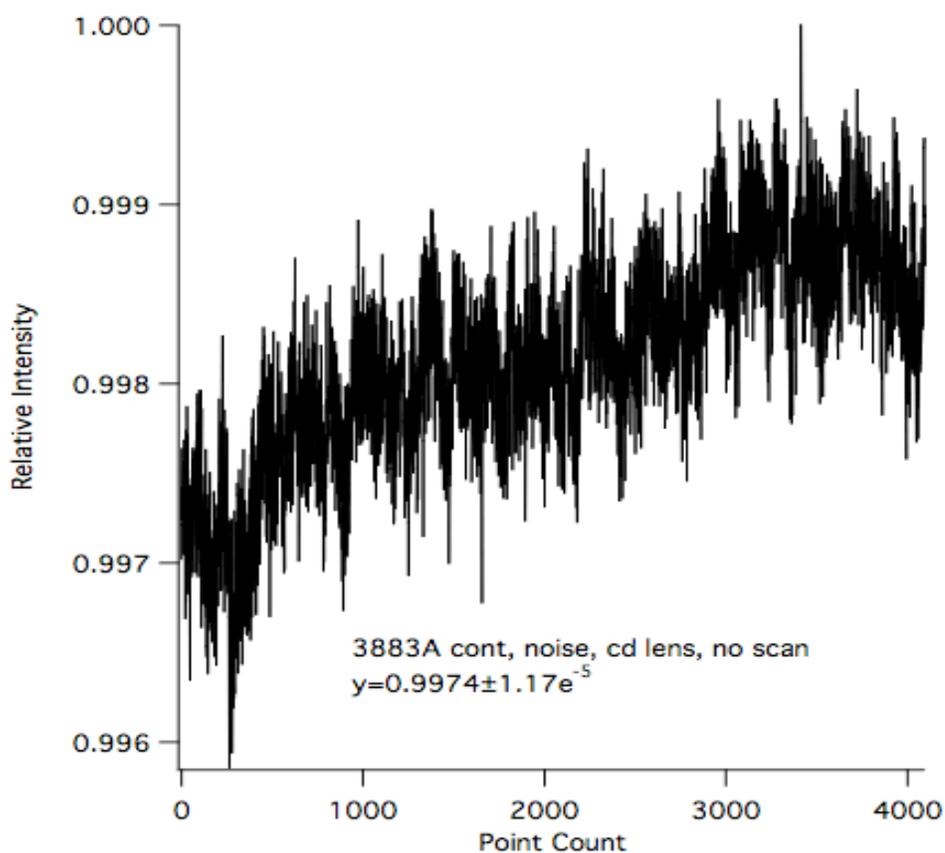

Figure 1. Example of system noise recorded at 3883 Å. The intensity (y) is normalized to 1.0.

Table 1. Results from system noise measurements.

| Wavelength (Å) | Linear fit (±1 σ) |
|---|---|
| 3129 | $y=0.995\pm2.86\times10^{-5}$ |
| 3883 | $y=0.997\pm1.17\times10^{-5}$ |
| 15648 | $y=0.999\pm1.31\times10^{-5}$ |
| 46880 | $y=0.993\pm7.79\times10^{-5}$ |

Crucial to all observations is a clear, cloud free sky. Any records interrupted by clouds have been deleted. An unexpected problem has been birds flying around the heliostat and through the beam. Such "glitches" have also been deleted.

An important aspect of this report is the 0.01% precision attained. A number of factors combine to make this possible. In arbitrary order of significance these are:

- Entrance slit 1.19×23.7 arc-sec (*i.e.* many granules averaged perpendicular to the scan, but good resolution in the scan direction)
- Single element detector (compared to a CCD array with its "flat fielding" errors)
- No optical interference fringing (fixed wavelength)
- Zero offset removal by the shutter under the slit
- 16 bit AD converter (1:65536)
- Multiple integrations (about 35 minutes)
- Good sky
- Large telescope aperture (suppresses scintillation)
- Large telescope aperture (low photon noise)
- Lens to average over the granular field (in noise tests)

Various aspects of system noise that are relevant here are discussed in Birney, Gonsalez, and Oesper (2006).

## 3. Results for scan observations

Examples of scans are given in Figure 2. Here we have superimposed the averaged scans at 3129, 4615, 5394, 15648, and 46700 Å. In Table 2 we list all the observations and include the one σ error from a linear fit across the mid 500 steps (267 arc-sec). The "file name" refers to a Virtual Observatory file (mentioned in Section 1), which the reader can access. These files are for the averaged limb-to-limb scans. Figure 3 shows an example of data for 3129 Å. The plot shows the middle 267 arc-sec of the N - S scan with a linear fit to it. How real are the features in Figure 3? To judge this we compare, in Figure 4, the temporally adjacent S – N scan also at 3129 Å. This ought to be the wavelength with the greatest granular contrast. The time difference is about 1 minute (depending on position in the scan). Recall that the slit length is 25 arc-sec, so we average over many granules. This shows that the features are real (*i.e.* solar), but changing somewhat with time.

We have also compared many N - S against W - E plots. An example is shown in Figure 5. No systematic differences are found.

Scattered light has not been taken into account. It decreases with wavelength from 0.01 at 3129 Å to 0.00015 at 46700 Å. These values can be read from the off-limb intensity files listed in Table 2.

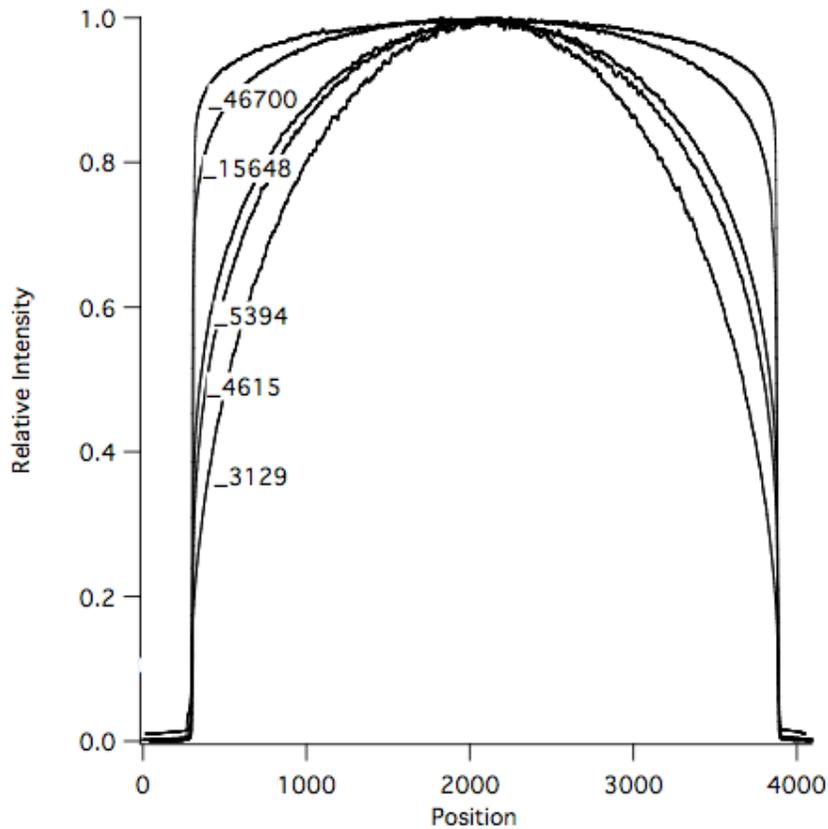

Figure 2. Samples of averaged scans at 3129, 4615, 5394, 15648, and 46700 Å. Position is in record units (1 – 4096), at 0.534 arc-sec per step. Because the start position varies slightly, each scan has been adjusted here so the scan records overlap.

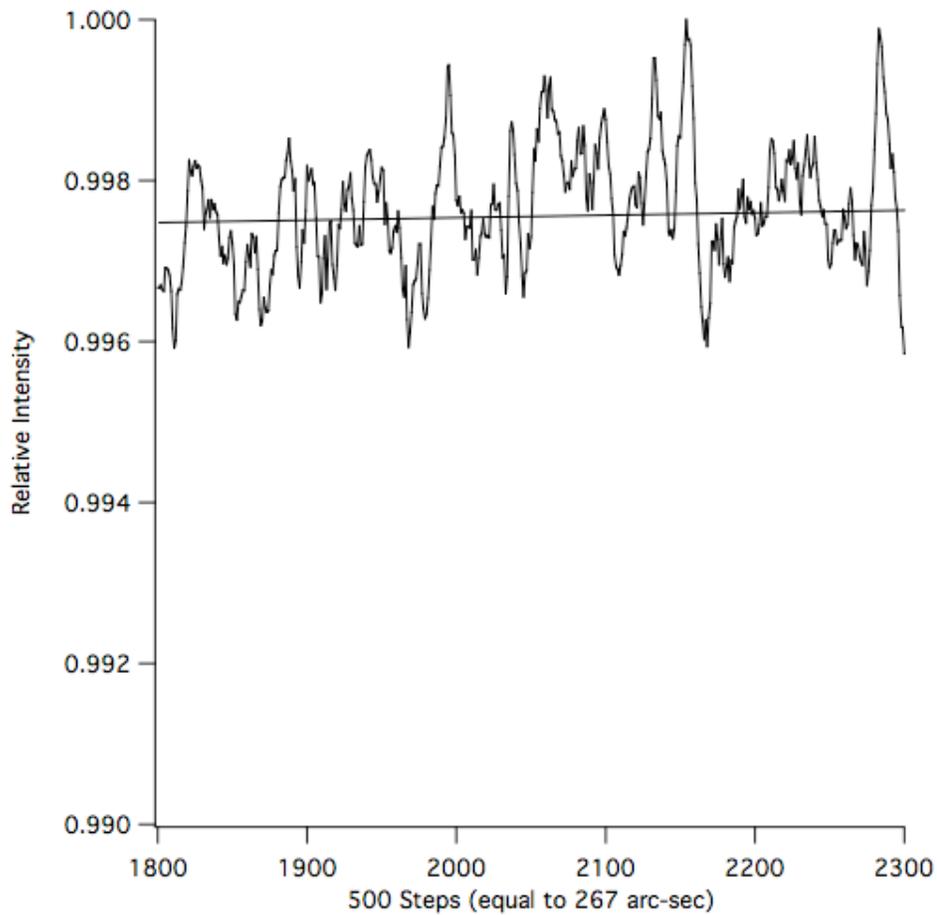

Figure 3. Expanded approximate mid-disk (267 arc-sec) portion of 3129 Å N - S scan. Again the exact position on the disk is not known (nor important for this demonstration).

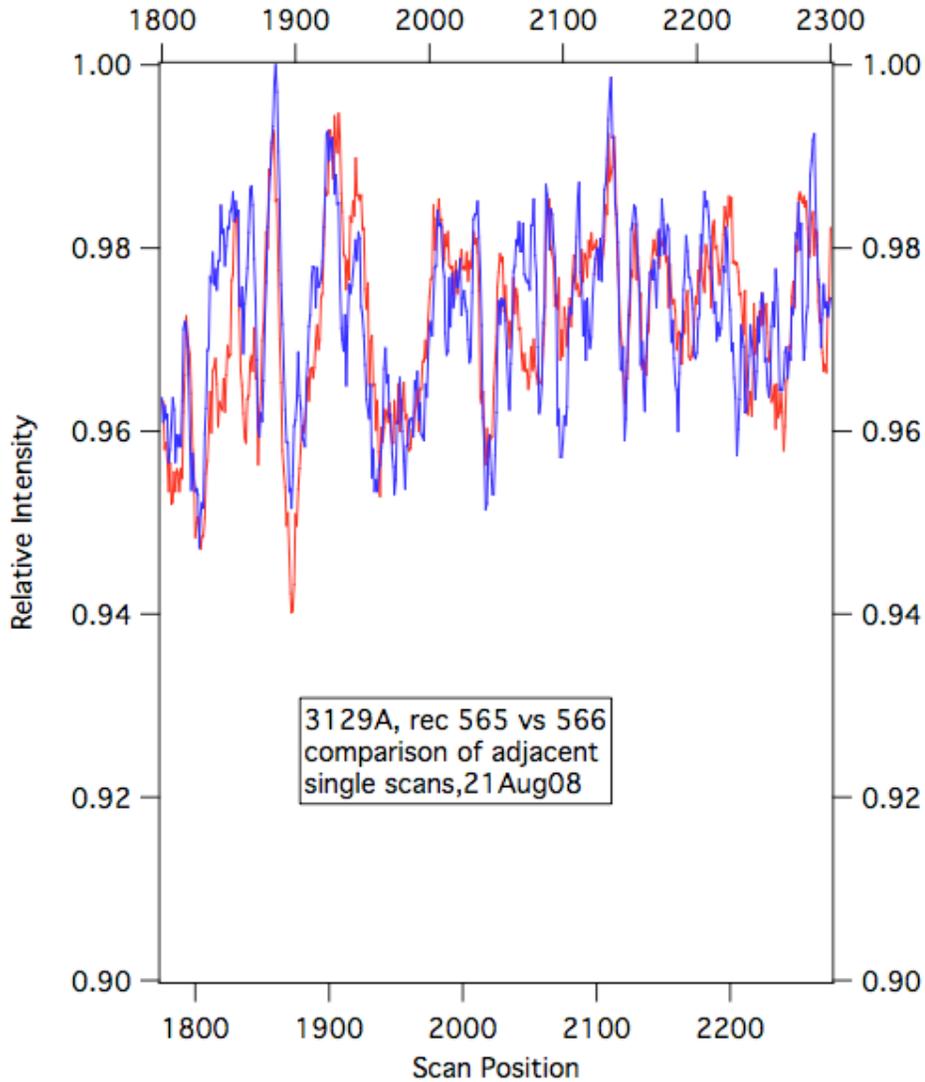

Figure 4. Two temporally adjacent scans at 3129 Å; same as Figure 3 with the following S-N adjusted in scan position to overlap in structure. This shows that larger features persist but details (granules or seeing?) change over the one minute lapsed time.

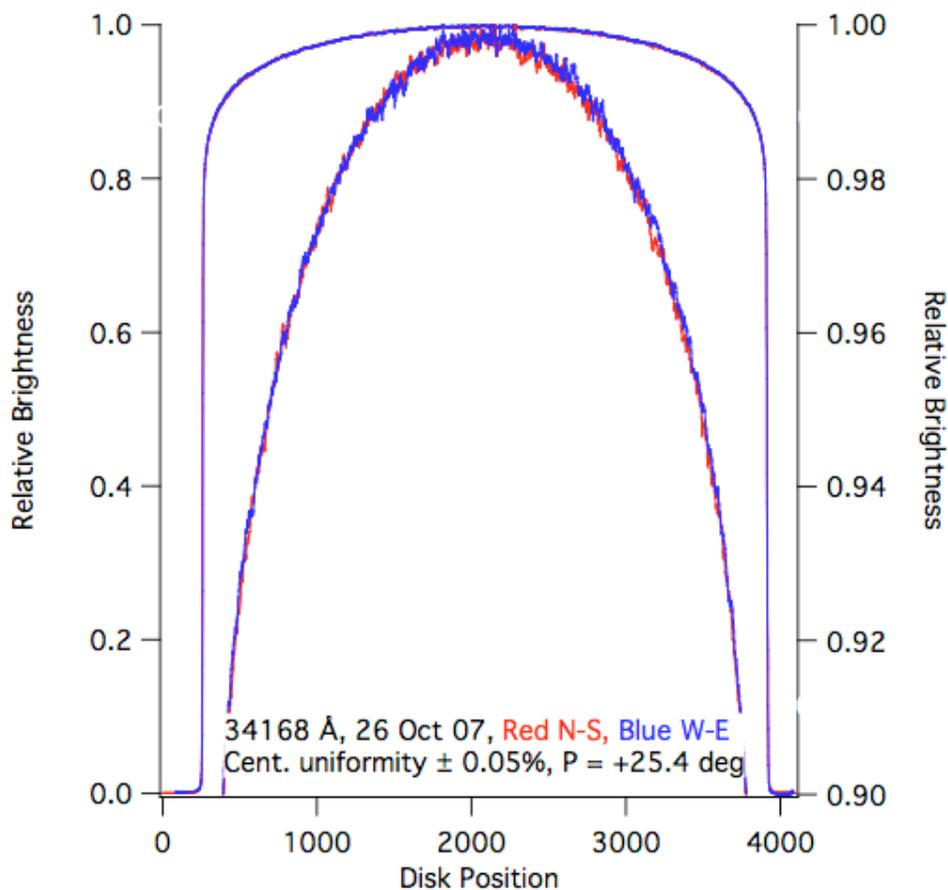

Figure 5 Comparison of entire disk normalized equatorial vs meridional scans at 34168 Å. Note that P = +25.4 deg so this is offset from heliographic coordinates. Plot full size and x10.

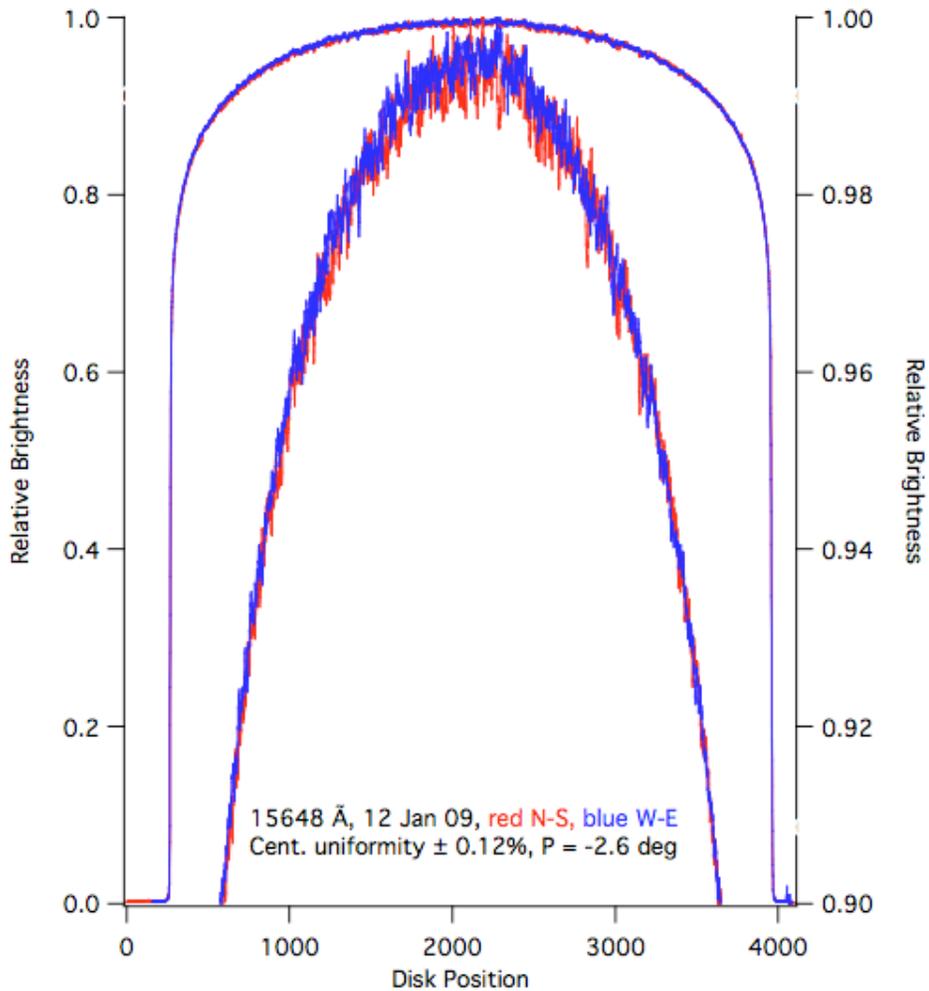

Figure 6. Comparison of near solar equatorial vs meridional scans at 15648 Å. As mentioned, this condition occurs briefly once a year around 7 July and 6 January. Plot full size and x10.

Table 2. Fluctuations in mid-limb-darkening scans from Oct 2007 to Jun 2008.

| Wave-Length Å | Date: | Scan | P-angle (deg) | Fluctuation mid-scan (1 σ) | File name (ftp site) |
|---|---|---|---|---|---|
| 3129 | 2008Mar25 | N-S | -25 | .0028 | 3129.205 |
| 3129 | 2008Mar28 | N-S | -25 | .0029 | 3129.793 |
| 3129 | 2008Aug21 | N-S | +18 | .0030 | 3129.565 |
| 3129 | 2008Aug21 | W-E | +18 | .0031 | 3129.585 |
| 3883 | 2008Apr14 | N-S | -26 | .0022 | 3883.405 |
| 4615 | 2008Jun03 | N-S | -15 | .0017 | 4615.217 |
| 4615 | 2008Jun03 | W-E | -15 | .0015 | 4615.239 |
| 5394 | 2008Jan21 | N-S | -7 | .0015 | 5394.393 |
| 8558 | 2008Jan19 | N-S | -6 | .0011 | 8558.77 |
| 15648 | 2007Oct23 | N-S | +25 | .00088 | 15648.523 |
| 15648 | 2007Oct24 | N-S | +25 | .0011 | 15648.633 |
| 15648 | 2007Oct24 | W-E | +25 | .00085 | 15648.653 |
| 15648 | 2007Oct25 | N-S | +25 | .0010 | 15648.815 |
| 15648 | 2007Oct25 | W-E | +25 | .0013 | 15648.835 |
| 15648 | 2007Oct26 | N-S | +25 | .0010 | 15648.1025 |
| 15648 | 2007Oct26 | W-E | +25 | .0010 | 15648.1045 |
| 15648 | 2009Jan12 | N-S | -2.6 | .0012 | 15648.281 |
| 15648 | 2009Jan12 | W-E | -2.6 | .0012 | 15648.293 |
| 34100 | 2007Oct25 | W-E | +25 | .00050 | 34100.859 |
| 34100 | 2007Oct25 | N-S | +25 | .00055 | 34100.879 |
| 34100 | 2007Oct26 | W-E | +25 | .00048 | 34100.1065 |

| 34100 | 2007Oct26 | N-S | +25 | .00047 | 34100.1085 |
| 46700 | 2008Apr16 | N-S | -26 | .00062 | 46700.679 |
| 46700 | 2008Apr16 | W-E | -26 | .00060 | 46700.697 |

To provide a measure of comparison between the observations and some standard solar models we have calculated the limb darkening curves for two standard empirically derived solar models shown in Table 3.

Table 3
HSRA    Gingerich, et. al. 1971
H-M     Holweger and Müller 1974

The emergent continuum radiation was calculated for each model using an LTE plane parallel formalism with the equation of state and opacity routines as described in Auer, Heasley, and Milkey (1972). These opacities should be quite adequate except for the violet end of the spectrum. The equation of transfer was solved using the standard Feautrier (1964) difference-equation method.

The resulting intensities were then normalized to the disk center intensity ($I/I_0$) for comparison with the observations. The abscissa is the usual $\mu = \cos\theta$. In evaluating these comparisons, one should remember that a uniform, plane-parallel, model is a crude approximation to the real conditions of the solar photosphere and limb darkening is only one of many solar observations that such models are asked to approximate.

We display as Figures 7, 8, 9, and 10 this comparison of models with disk center observations out to ~ $\mu = 0.1$ ($\mu=0.0$ at the limb; $\mu=1.00$ at disk center). In these plots we have averaged the two halves of the observations after locating the mid-position. Our shortest wavelength continuum cannot be considered unblended, although this position, 3129.6 Å, is the high point in the local spectrum (see Houtgast, 1970, Wallace, Hinkle, and Livingston, 2007). As a result of this unknown blending, the comparisons with model atmospheres are compromised by the limitation to only continuum opacity sources in the intensity calculation.

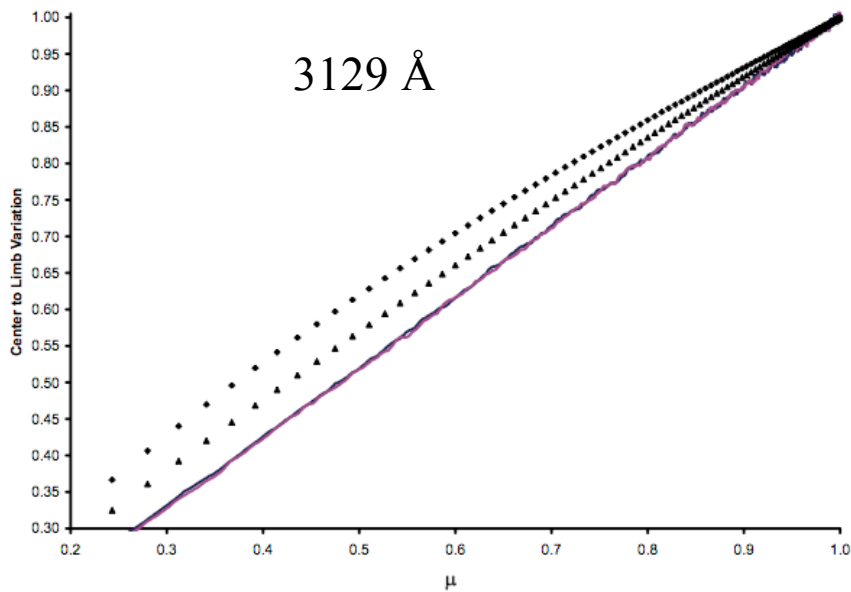

Figure 7. Comparison of observed intensity at 3129 Å N-S (purple), W-E (blue), and the models HSRA (triangle), H-M (diamond), and RadEq (square).

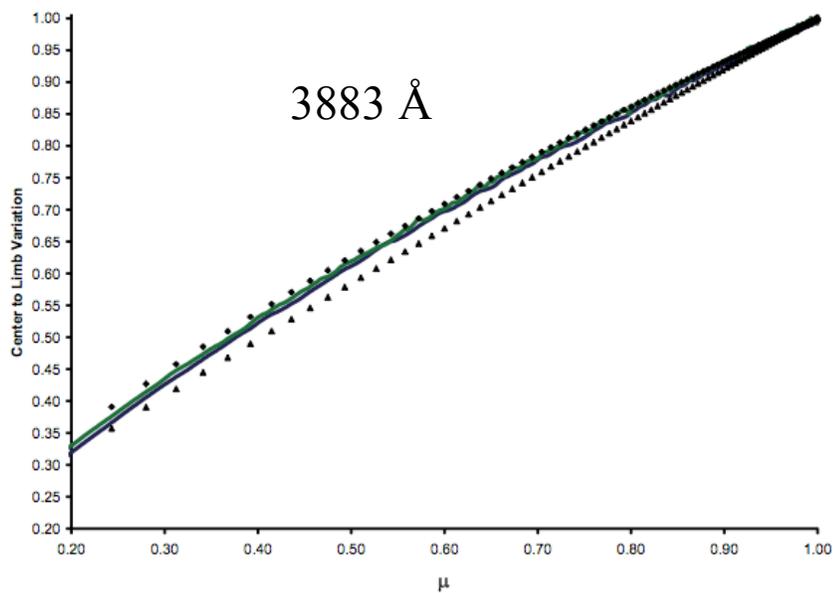

Figure 8. Comparison of 3883 Å N-S (blue) and W-E (green) with HSRA, H-M, and

RadEq.

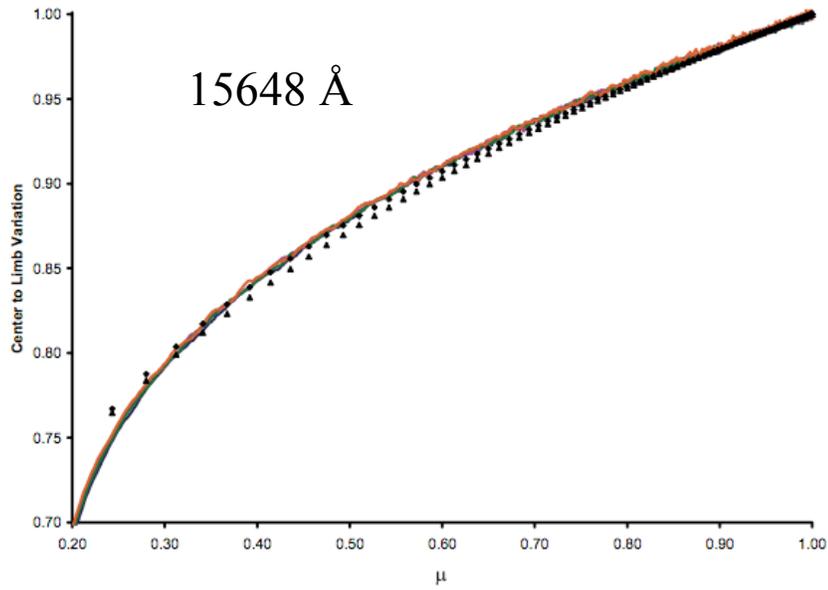

Figure 9. Comparison of opacity minimum 15648 Å N-S (blue, red), W-E (green, orange), with HSRA, H-M and

RadEq.

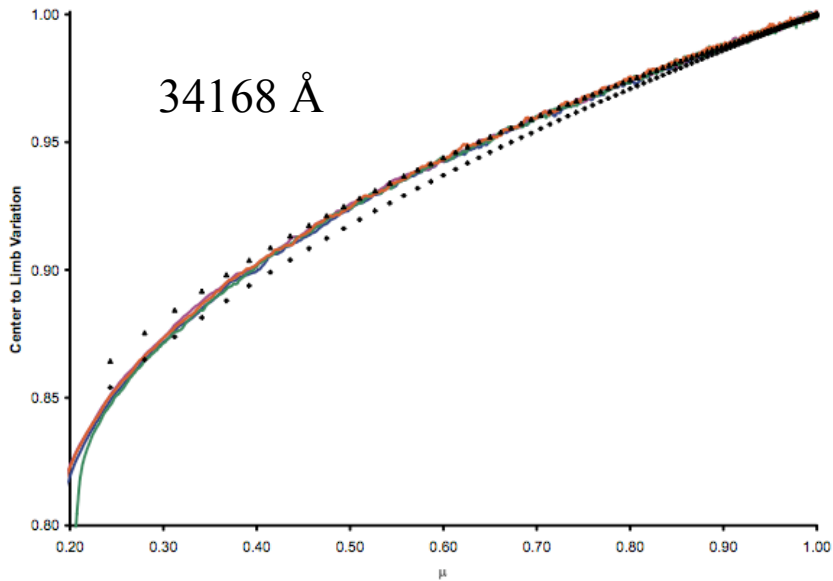

Figure 10. Comparison of 34168 Å N-S (blue, orange) and W-E (green, red) scans with HSRA, H-M, and RadEq models.

## 4. Discussion and Conclusions

We have made scans in both equatorial and meridional directions. System noise is under 0.01 % for the circumstances of the observations. Solar fluctuations near disk center decrease with wavelength from 0.3% at 3129 Å to 0.06 % at 46700 Å. The latter value corresponds to 3.1 K assuming the Rayleigh-Jeans distribution. These values are much less than examples given by Pierce *et al.* (1977), presumably because of our quiet Sun and averaging process. These residuals could arise from the unresolved and/or the long-term contributions of granulation. Granulation signals would be expected to decrease with increase of wavelength (Muller, 1988). Brandt and Getling (2004) have noted that granulation patterns can have a lifetime much greater than normally expected (Muller, 1988). Other fluctuation sources, such as *p*-mode oscillations, are unlikely (see Table II in Jimenez *et al.*, 1990).

As shown in Table 2, there is no evidence for differences in the equator *vs.* the meridian; see also Figure 5. This does not support the finding of Rast, Ortiz, and Meisner (2008), although our time and spatial coverage differ from theirs. Remember also that, because of sky coordinates, we do not usually sample the true polar regions. The agreement with model atmospheres is considered satisfactory, at least near disk center. We have not studied the extreme limb in this paper.

In summary, the quiet solar disk is observed to be uniform in continuum limb darkening within 0.3% at 3129 Å to .05% at 34100 Å. These results could serve as reference for the detection of exoplanets on space experiments such as the Kepler Mission and COROT.

## Acknowledgements

We are indebted to Jim Brault for his development in the 1960s of the unique and powerful photometry system at the

McMath-Pierce Telescope (Brault, *et al*. 1971). Brault's awareness of "1/*f* noise", where *f* is frequency, led to rapid-scan data taking with automatic DC corrections. This technique is crucial to the present work. Dick Joyce installed a new InSb detector in the "Babo" dewar. Claude Plymate designed and built the electronic filter. Jim Heasley and Han Uitenbroek gave advice on modeling. Frank Hill was consulted on the possible role of solar oscillations. John Britanic has recently upgraded photometry to eliminate certain obsolete hardware components, like 9-track tapes. Serge Koutchmy called our attention to earlier limb-darkening literature.